\documentclass[twoside,reqno]{HERON}
\usepackage{epsfig,cite,colordvi}
\usepackage{graphicx}
\usepackage{url}
\usepackage{amssymb,amsmath,amscd,epsf}
\usepackage{times}
\usepackage{makeidx}
\usepackage{amsfonts}

\renewcommand\bar\overline
\newcommand\gO{\text{O}}
\newcommand\SO{\text{SO}}
\newcommand\lo{\mathfrak{o}}
\newcommand\Cl{\text{Cl}}
\newcommand\tr{\text{tr}\,}
\renewcommand\log{\text{log}\,}
\newcommand\pf{\text{pf}}

\begin{document}

\title{``Onishi'' formulas}

\runningheads{``Onishi'' formulas}{K. Neerg\aa rd}

\begin{start}

\author{K. Neerg\aa rd}{1}

\index{Neerg\aa rd, K.}

\address{Fjordtoften 17, 4700 N\ae stved, Denmark}{1}

\begin{Abstract}
  Among the linear transformations of the span of annihilation and
  creation operators pertaining to a finite-dimensional space of
  single-fermion states, Bogolyubov transformations are those which
  preserve the anticommutator. Since this is a symmetric bilinear
  form, Bogolyubov transformations are orthogonal. A Bogolyubov
  transformation defines a quasi-fermion vacuum killed by the
  transformed annihilation operators. Applying the generator
  coordinate method to quasi-fermion vacua, including projecting a
  quasi-fermion vacuum onto the eigenspace of conserved quantum
  numbers, requires calculating the overlap amplitude between
  different vacua. Several different formulas for this amplitude in
  terms of the parameters of the generating Bogolyubov transformations
  carry in the literature the name of an ``Onishi'' formula, referring
  to a 1966 paper by Onishi and Yoshida. In particular the formula
  called so in the much cited book by Ring and Schuck differs from
  that of Onishi and Yoshida and has a wider scope. These formulas are
  often written with a square root and thus have a sign ambiguity. I
  show that this sign ambiguity stems from the fact that the
  representation of the orthogonal group of Bogolyubov transformations
  on the space of any number of fermions inhabiting the single-fermion
  state space is its spin representation, which is double valued. In
  the case of the general formula of Ring and Schuck, the sign
  ambiguity is therefore unavoidable. The formula of Onishi and
  Yoshida only applies when both vacua have non-zero overlap with the
  physical vacuum. They are then generated by a simply connected
  subset of the group. This renders their phases well defined, so the
  formula of Onishi and Yoshida has no sign ambiguity.

  In 1983, W\"ust and I devised a method to determine the unambiguous
  sign of the square root in the formula of Onishi and Yoshida. A
  different method was proposed by Robledo in 2009. I discuss the
  relation between these methods and show, in particular, how
  Robledo's formula follows directly from that of Onishi and Yoshida.
  Notably, neither Onishi and Yoshida nor their contemporary authors
  ever presented a derivation of their formula. Robledo based the
  derivation of his formula on Berezin integration. I discuss various
  ways to derive the formula of Onishi and Yoshida and various
  alternative ways to derive Robledo's formula.
\end{Abstract}

\end{start}

\section{\label{sec:intr}Introduction}

The term ``Onishi'' formula, introduced in~\cite{ref:Rin80}, refers to
a family of related formulas for the overlap amplitude of two
Bogolyubov quasi-nucleon (more generally quasi-fermion) vacua. As a
common feature, these formulas display a square root, which gives rise
to an apparent sign ambiguity, first discussed by W\"ust and
me~\cite{ref:Nee83} and referred to in the literature as ``the sign
problem of the Onishi formula'' or the like~\cite{ref:Sch04,
  ref:Oi05,ref:Ben09,ref:Rob09,ref:Ave12,ref:Ber12,ref:Miz12,ref:Miz18,
  ref:Bal18,ref:Por22}. It is shown below that for some members of the
family, this sign ambiguity is real and unavoidable while in some
other cases it is not. The basis for my analysis is the observation
that the group of Bogolyubov transformations pertaining to a given
single-fermion state space is isomorphic to an orthogonal group and
the representation of this group on the Fock space, the space of
states of any number of fermions from the single-fermion state space,
is just its so-called spin representation, described by Brauer and
Weyl in 1935~\cite{ref:Bra35}. It is the fundamental double-valuedness
of the spin representation that gives rise to the sign ambiguity of
some Onishi formulas.

I explain in Sec.~\ref{sec:ort} the said isomorphism between a group of
Bogolyubov transformations and an orthogonal group. I then introduce
the spin representation and explain its double-valuedness. I also
introduce a coordinate representation, which displays in matrix form
the action of the Bogolyubov transformation on annihilation and
creation operators. It is pointed out that the double-valuedness of
the spin representation gives rise to a fundamental sign ambiguity in
the representation of Bogolyubov transformations by operators on the
Fock space. Infinitesimal Bogolyubov transformations are discussed and
properties of their coordinate and spin representations presented,
leading to an explicit expression for the spin representation image of
an arbitrary Bogolyubov transformation in the physically important
case when the equivalent orthogonal transformation is unitary and
proper.

Quasi-fermion vacua are introduced in Sec.~\ref{sec:vac} and it is
explained that the sign ambiguity of the Fock space representation
translates to fundamental sign ambiguity in the assignment of a
quasi-fermion vacuum to a given Bogolyubov transformation. I then
consider overlap amplitudes between different quasi-fermion vacua.
Restricting my scope to the case of unitary and proper Bogolyubov
transformations and even-dimensional single-fermion state spaces and
using the expression for the Fock space representations of such
Bogolyubov transformations obtained in Sec.~\ref{sec:ort}, I derive
the formula for the overlap amplitude that is called the ``Onishi''
formula in~\cite{ref:Rin80} under more general assumptions than made
there.

The formula referred to by this name and presented originally by
Onishi and Yoshida~\cite{ref:Oni66} is different. At the cost of a
more limited scope it has \emph{no} sign ambiguity. This is explained
in Sec.~\ref{sec:Oni}. Variants of the formula of Onishi and Yoshida
are discussed. Analysing the formula is shown, in particular, to lead
to a general theorem on the characteristic roots of a product of two
skew symmetric matrices which follows more directly from the
properties of the Pfaffian of a skew symmetric matrix. As discussed in
section~\ref{sec:Rob}, this provides a direct link between the
original formula of Onishi and Yoshida and a variant in terms of the
Pfaffian presented by Robledo in 2009. The present paper is summarised
in Sec.~\ref{sec:sum}.

\section{\label{sec:ort}Bogolyubov transformations are orthogonal}

Consider a space of single-fermion states of finite dimension $d$ with
orthonormal basic states $| i \rangle$ and corresponding annihilation
operators $a_i$ and creation operators $a_i^\dagger$. These operators
span the space $\mathcal F$ of \textit{field operators}
$\alpha,\beta,\dots$ and generate the Clifford algebra $\Cl(2d)$. A
\textit{Bogolyubov transformation} $g$ is a linear transformation of
$\mathcal F$ that preserves the anticommutator $\{\alpha,\beta\}$.
Since this is a non-degenerate, symmetric bilinear form, the group of
Bogolyubov transformations can be identified with the group $\gO(2d)$
of orthogonal transformations in $2d$ complex dimensions. The
(faithful) \textit{coordinate representation} $g \mapsto G$ of
$\gO(2d)$, where $G$ is a $2d \times 2d$ matrix, is defined by
$g \alpha_- = \alpha_- G$, where $\alpha_-$ denotes a row of basic
field operators. A column of such operators is denoted by $\alpha_|$.
Usually a Bogolyubov transformation is assumed to be also
\textit{unitary}. It then preserves also the Hermitian inner product
$\{\alpha^\dagger,\beta\}$. The group of unitary Bogolyubov
transformations form the subgroup $\gO(2d,\mathbb R)$ of orthogonal
transformations in $2d$ real dimensions. One can distinguish finally
between \textit{proper} Bogolyubov transformations $g$ with
$\det g = 1$ and \textit{improper} Bogolyubov transformations $g$ with
$\det g = -1$. The proper Bogolyubov transformations form subgroups
$\SO(2d)$ and $\SO(2d,\mathbb R)$ of $\gO(2d)$ and
$\gO(2d,\mathbb R)$, respectively. In the Fock space representation to
be discussed below, the improper Bogolyubov transformations change the
number parity.

Brauer and Weyl showed that $\gO(2d)$ has a \emph{double-valued}
representation $g \mapsto \bar g$ in $\Cl(2d)$~\cite{ref:Bra35}, known
as its \textit{spin representation}. That the representation is
double-valued means that $g_1 g_2 \mapsto \pm \bar g_1 \bar g_2$ and
\emph{a continuous path in \textup{$\gO(2d)$} (indeed in
  \textup{$\SO(2d,\mathbb R)$}) from the identity $1$ back to itself
  takes the representation image from $1$ to $-1$.} The representation
image $\bar g$ is defined by (i) $(g \alpha) \bar g = \bar g \alpha$
and (ii) $(\tau \bar g) \bar g = 1$, where the operator $\tau$ inverts
the factor order in a product of field operators. Here (i) determines
$\bar g$ up to a numeric factor and (ii) fixes this factor within a
sign. In consequence of the double-valuedness of the spin
representation it is \emph{impossible to map the entire group of
  Bogolyubov transformations continuously and single-valuedly into
  \textup{$\Cl(2d)$} in such a way that this map $g \mapsto \bar g$
  obeys \textup{(i)} for every $g$ and $\alpha$.} Indeed, even
relaxing the normalisation condition (ii) by the multiplication of
$\bar g$ by a continuous numeric factor cannot remove the
double-valuedness. This holds upon restriction to $\gO(2d,\mathbb R)$,
$\SO(2d)$ or $\SO(2d,\mathbb R)$ and will be seen to be the origin of
the sign ambiguity mentioned in the introduction.

\textit{Infinitesimal} Bogolyubov transformations $x$ obey
$\{x \alpha , \beta\} + \{\alpha , x \beta\} = 0$ and form
the Lie algebra $\lo(2d)$. In the basis of Hermitian field operators
$a_i + a^\dagger_i$ and $-i (a_i - a^\dagger_i)$, the coordinate
representation image of $x$ is a skew symmetric matrix $X$. Its spin
representation image $\bar x$ is $\tfrac12 \alpha_- X \alpha_|$. In
the more usual basis $a_1,\dots,a_d,a_1^\dagger,\dots,a_d^\dagger$,
one gets
\begin{equation}\label{eq:x-o2dR}
  X \begin{pmatrix}0&1\\1&0\end{pmatrix}
  = - \begin{pmatrix}0&1\\1&0\end{pmatrix} X^T , \quad
 \bar x =
    \tfrac12 \alpha_- X \begin{pmatrix}0&1\\1&0\end{pmatrix} \alpha_|
\end{equation}
in terms of matrices with $d \times d$ blocks. (Details
in~\cite{ref:Nee23}.)

In most applications in nuclear physics, $g \in \SO(2d,\mathbb R)$,
that is, $g$ is \emph{unitary} and \emph{proper}. Then in the basis of
Hermitian field operators, $G$ is real orthogonal and thus
real-orthogonal equivalent to a block diagonal matrix with diagonal
blocks
\begin{equation}
  \begin{pmatrix}\cos\phi&\sin\phi\\-\sin\phi&\cos\phi\end{pmatrix}
  = \exp \begin{pmatrix}0&\phi\\-\phi&0\end{pmatrix}
\end{equation}
with real $\phi$. Hence $G = \exp X$, where $X$ is skew symmetric.
Then $X$ represents an infinitesimal Bogolyubov transformation $x$,
and $g = \exp x$. In the basis of annihilation and creation operators,
one gets from~\eqref{eq:x-o2dR} that~\cite{ref:Nee83}
\begin{equation}\label{eq:gbar}
 \bar g = \exp \bar x = \exp \tfrac12
   \alpha_- X \begin{pmatrix}0&1\\1&0\end{pmatrix} \alpha_| .
\end{equation}
The angles $\phi$ are determined by $G$ only modulo $2 \pi$ and adding
$2 \pi$ to a single $\phi$ gives rise in~\eqref{eq:gbar} exactly to a
change of sign of $\bar g$. This demonstrates the double-valuedness
explicitly.

\section{\label{sec:vac}Quasi-fermion vacua and their overlaps}

The field operators act on the space $\mathcal K$ of states of any
number of fermions from the single-fermion state space. Formally,
$\mathcal K = \Cl(2d) |\rangle$, where $|\rangle$ is a \textit{vacuum
  state} characterised by $a_i |\rangle = 0$ for every $i$. In the
terminology of Brauer and Weyl, it is the \textit{spinor space}
associated with $\Cl(2d)$. Given a Bogolyubov transformation $g$, a
\textit{quasi-fermion vacuum} $|g\rangle = 0$ obeys
$g a_i |g\rangle = 0$ for every $i$. Part (i) of the definition of the
spin representation implies $|g\rangle \propto \bar g |\rangle$, so
the double-valuedness of the spin representation makes it
\emph{impossible to map the entire group of Bogolyubov transformations
  continuously and single-valuedly into $\mathcal K$ in such a way
  that this map $g \mapsto |g\rangle$ obeys $g a_i |g\rangle = 0$ for
  every $g$ and $i$}. It is convenient to set $\langle|\rangle = 1$
and $|g\rangle = \pm \bar g |\rangle$. Then $\langle g|g\rangle = 1$
when $g$ is unitary, and $|g_1g_2\rangle = \pm \bar g_1 |g_2\rangle$.

Structure calculations such as the projection of quasi-fermion vacua
onto the state space of some conserved quantum numbers requires the
calculation of overlap amplitudes $\langle g_1 | g_2 \rangle$. I
assume from now on that $g_1$ and $g_2$ are unitary, which case covers
the applications in nuclear physics. Then by
$\langle g_1 | g_2 \rangle = \pm \langle|\bar g_1^\dagger \bar g_2
|\rangle = \pm \langle|\bar g_1^{-1} \bar g_2 |\rangle = \pm
\langle|\bar {g_1^{-1} g_2} |\rangle$, an expression for
$\langle|\bar g |\rangle$ for arbitrary $g$ suffices for the
calculation up to a sign of every $\langle g_1 | g_2 \rangle$. It can
be shown that for improper $g$, the spin representation image $\bar g$
changes the number parity~\cite{ref:Nee23}. Since the vacuum has even
number parity, $\langle|\bar g |\rangle = 0$ in this case. I can
therefore assume further that $g$ is proper, that is,
$g \in \SO(2d,\mathbb R)$. If also $d$ is even, as usual in nuclear
structure applications, one can then apply the Block-Messiah
decomposition~\cite{ref:Blo62}
\begin{equation}\label{eq:Bloch}
  G = \begin{pmatrix} D^* & 0 \\ 0 & D \end{pmatrix}
    \begin{pmatrix} U & V \\ V & U \end{pmatrix}
    \begin{pmatrix} C^* & 0 \\ 0 & C \end{pmatrix}
\end{equation}
to the coordinate representation image $G$ in the basis of
annihilation and creation operators. Here $C$ and $D$ are unitary and
$U$ and $V$ are block diagonal with diagonal blocks
\begin{equation}
  \begin{pmatrix} u & 0 \\ 0 & u \end{pmatrix}
  \quad \text{and} \quad
  \begin{pmatrix} 0 & v \\ -v & 0 \end{pmatrix} ,
\end{equation}
where $u,v \ge 0$ and $u^2 + v^2 = 1$. Since each factor in
\eqref{eq:Bloch} is the coordinate representation image of some
$g \in \SO(2d,\mathbb R)$, the Block-Messiah decomposition corresponds
to a decomposition $g = g_D g_W g_C$ with
$g_D,g_W,g_C \in \SO(2d,\mathbb R)$.

The contribution to $\langle|\bar g |\rangle$ of each factor $g_D$,
$g_W$ and $g_C$ can be evaluated by means of \eqref{eq:gbar}. The
result is $\langle| \, \bar g_D = \sqrt{|D^*|} \, \langle|$ and
$\bar g_C \, | \rangle = | \rangle \, \sqrt{|C^*|}$ with indefinite
signs due to the multi-valuedness of each $X$ as a function of the
corresponding $G$, and $\langle | \bar g_W | \rangle = \prod u
= \sqrt{|U|} \ge 0$~\cite{ref:Nee23}. When $G$ is expressed in the
form
\begin{equation}
  G = \begin{pmatrix} A^* & B \\ B^* & A \end{pmatrix} ,
\end{equation}
this combines to
$\langle | \bar g | \rangle = \sqrt{|D^* U C^*|} = \sqrt{|A^*|}$ with
an indefinite sign of the square root as anticipated from the
indefinite sign of $\bar g$. In the present derivation, the
indefiniteness is seen to be due to indefinite signs of $\sqrt{|D^*|}$
and $\sqrt{|C^*|}$, while $\sqrt{|U|} \ge 0$. The equivalent result
$\langle g | \rangle = \sqrt{|A|}$ is derived in~\cite{ref:Rin80} in
the case $D = C = 1$, which can be generalised to the case when $A$ is
Hermitian and positive semi-semidefinite ($D = C^\dagger$). In that
case, $|A^*| = |A| \ge 0$. The expression
$\langle g | \rangle = \sqrt{|A|}$ is called the ``Onishi'' formula
in~\cite{ref:Rin80}.

\section{\label{sec:Oni}Formula of Onishi and Yoshida}

The name of an ``Onishi'' formula refers to a 1966 paper by Onishi and
Yoshida~\cite{ref:Oni66}. These authors consider, however, not the
normalised state $|g\rangle$, but the unnormalised state\sloppy
\begin{equation}\label{eq:gtilde}
  | \tilde g \rangle = \dfrac {| g \rangle} {\langle | g \rangle}
  = \exp \tfrac12 a^\dagger_- F a^\dagger_| \, | \rangle ,
\end{equation}
where $F = ( B A^{-1} )^* = - F^T$. The last expression
in~\eqref{eq:gtilde} is the Thouless expansion~\cite{ref:Tho60}. Note
that $| \tilde g \rangle$ has \emph{no phase ambiguity}. Its
normalisation is fixed by $\langle | \tilde g \rangle = 1$. This comes
at the cost that it is defined only when $\langle | g \rangle \ne 0$
or, equivalently, $|A| \ne 0$. Onishi and Yoshida write
\begin{equation}
  \langle \tilde g_1 | \tilde g _2 \rangle
  = \exp \tfrac12 \tr \log ( 1 + F_1^\dagger F_2 ) .
\end{equation}
Notably, a \emph{derivation} of this expression is given by neither
Onishi and Yoshida nor their contemporary
authors~\cite{ref:Bec70,ref:Goe72,ref:Goe73}, all of whom refer
to~\cite{ref:Oni66}. A fairly short derivation is shown
in~\cite{ref:Nee23}, and other derivations were also presented more
recently~\cite{ref:Miz18,ref:Por22}.

Note that $\exp \tfrac12 a^\dagger_- F a^\dagger_|$ is a
\emph{polynomial} in the entries of $F$ because at most $d$ creation
operators can be multiplied together to a non-zero result. Therefore
in
\begin{equation}\label{eq:g1g2z}
  \langle \tilde g_1 | \tilde g _2(z) \rangle
  = \exp \tfrac12 \tr \log ( 1 + z F_1^\dagger F_2 )
  = \sqrt { | 1 + z F_1^\dagger F_2 | } ,
\end{equation}
where
$\tilde g(z) = \exp \tfrac12 a^\dagger_- z F a^\dagger_| \, |\rangle$,
the first expression, and therefore also the second and the third, are
polynomials in $z$ provided in the latter two, we take $\log 1 = 0$
and $\sqrt{1} = 1$ and require continuity in $z$. It follows in
particular that then both the latter are well defined for every $z$.
The last expression in \eqref{eq:g1g2z} is a polynomial in $z$ only if
the non-zero characteristic roots of $F_1^\dagger F_2$ have even
multiplicities, so we can write, in fact,
\begin{equation}
  \langle \tilde g_1 | \tilde g _2 \rangle = \prod\nolimits' (1 + r) ,
\end{equation}
where the product runs over one out of each pair of equal
characteristic roots $r$ of $F_1^\dagger F_2$~\cite{ref:Nee83}.

The only assumption was that $F_1^\dagger$ and $F_2$ be skew
symmetric, so it was proved, in fact, that \textit{whenever two square
  matrices $P$ and $Q$ of equal dimensions are skew symmetric, the
  characteristic roots of their product $PQ$ have even
  multiplicities.} This follows much more directly from the following
result due to Cayley: \textit{When $P$ is skew symmetric,
  \textup{$|P| = (\pf \, P)^2$}, where \textup{$\pf \, P$} is a
  polynomial in the entries of $P$} (called its
\textit{Pfaffian})~\cite{ref:Cal49}. One calculates, in fact,
\begin{multline}\label{eq:|1+zPQ|}
  |1 + z P Q| = \begin{vmatrix}1 + z P Q & - z P\\ 0 & 1 \end{vmatrix}
  = \begin{vmatrix} z P & 1\\ -1 & Q\end{vmatrix}
      \begin{vmatrix} Q & -1 \\ 1 & 0\end{vmatrix}
  = \begin{vmatrix} z P & 1 \\ -1 & Q \end{vmatrix} \\
  = \left(\pf
    \begin{pmatrix} z P & 1 \\ -1 & Q \end{pmatrix} \right)^2 ,
\end{multline}
whence the assertion follows~\cite{ref:Nee23}. A more elaborate proof
under certain restricting assumptions is found in~\cite{ref:Oi19}.

\section{\label{sec:Rob}Robledo formula}

A standard expression for the Pfaffian~\cite{ref:Cal49} gives
\begin{equation}
  \pf \begin{pmatrix} 0 & 1 \\ -1 & Q \end{pmatrix}
  = (-1)^{d(d-1)/2} .
\end{equation}
Setting $\sqrt{1} = 1$ and assuming continuity in $z$, one can
therefore write \eqref{eq:|1+zPQ|} in the form
\begin{equation}
  \sqrt{|1 + zPQ|}  = (-)^{d(d-1)/2} \,
    \pf \begin{pmatrix} zP & 1 \\ -1 & Q \end{pmatrix} ,
\end{equation}
so that for $z = 1$, the expression \eqref{eq:g1g2z} becomes 
\begin{equation}
  \langle \tilde g_1 | \tilde g _2 \rangle = (-)^{d(d-1)/2} \,
    \pf \begin{pmatrix} F_1^\dagger & 1 \\ -1 & F_2 \end{pmatrix} .
\end{equation}
In 2009, Robledo got this expression by Berezin integration
\cite{ref:Ber66} in a slightly different but equivalent
form~\cite{ref:Rob09}. Other derivations have appeared more
recently~\cite{ref:Miz18,ref:Por22}, and yet another, combinatoric,
one is presented in~\cite{ref:Nee23}.

\section{\label{sec:sum}Summary}

The group of Bogolyubov transformation pertaining to a given
finite-dimensional single-fermion state space was explained to be
isomorphic to an orthogonal group and its representation on the Fock
space, which is the space of states of any number of fermions from the
single-fermion state space, to be just the spin representation of this
group. The double-valuedness of the spin representation was shown to
lead to a fundamental sign ambiguity in the assignment of an operator
on the Fock space to a given Bogolyubov transformation, which
translates to a corresponding sign ambiguity in the assignment of a
quasi-fermion vacuum to such a transformation and also to a sign
ambiguity in the formula for the overlap amplitude of two
quasi-fermion vacua that is called the ``Onishi'' formula in the much
cited book by Ring and Schuck~\cite{ref:Rin80}. On the other hand the
original formula of Onishi and Yoshida which is referred to by this
name has no sign ambiguity. This comes at the cost of a more limited
scope. Versions of this formula were discussed. Analysing it leads to
a general theorem on the characteristic roots of the product of two
skew symmetric matrices which follows more directly from the
properties of the Pfaffian of a skew symmetric matrix. This provides a
direct link between the original formula of Onishi and Yoshida and a
variant in terms of the Pfaffian presented by Robledo in 2009.

\bibliography{rila-paper}

\end{document}